\documentclass[prb,twocolumn,aps]{revtex4}
\usepackage{graphicx}

\begin{document}
\title{
rf SQUID metamaterials
}
\author{N. Lazarides
}
\affiliation{
Department of Physics, University of Crete,
and Institute of Electronic Structure and Laser,
Foundation for Research and Technology--Hellas,
P. O. Box 2208, 71003 Heraklion,  Greece, \\
and Department of Electrical Engineering,
Technological Educational Institute of Crete,
P. O. Box 140, Stavromenos, 71500, Heraklion, Crete, Greece
}
\author{G. P. Tsironis
}
\affiliation{
Facultat de Fisica,
Department d'Estructura i Constituents de la Materia,
Universitat de Barcelona, 
Av. Diagonal 647, E-08028 Barcelona, Spain, \\
and Department of Physics, University of Crete,
and Institute of Electronic Structure and Laser,
Foundation for Research and Technology--Hellas,
P. O. Box 2208, 71003 Heraklion,  Greece
}

\begin{abstract}
An rf superconducting quantum interference device (SQUID) array in an 
alternating magnetic field is investigated  
with respect to its effective magnetic permeability,
within the effective medium approximation.
This system acts as an inherently nonlinear magnetic metamaterial,
leading to negative magnetic response, and thus negative permeability,
above the resonance frequency of the individual SQUIDs.  
Moreover, the permeability exhibits oscillatory behavior at low field
intensities, allowing its tuning by a slight change of the intensity 
of the applied field.
\end{abstract}


\maketitle
An rf superconducting quantum interference device (SQUID) consists
of a superconducting ring interrupted by a Josephson junction (JJ).
\cite{Likharev}
When driven by an alternating magnetic field, the induced
supercurrents around the ring are determined by the JJ
through the celebrated Josephson relations.  This system exhibits
rich nonlinear behavior, including chaotic effects.
\cite{Fesser}
Recently, quantum rf SQUIDs have attracted great attention,
since they constitute essential elements for quantum computing.
\cite{Bocko}
In this direction, rf SQUIDs with one or more zero and/or $\pi$ 
ferromagnetic JJs have been constructed.\cite{Yamashita}
In this Letter we show that rf SQUIDs may serve as constitual elements 
for nonlinear 
magnetic metamaterials (MMs), i.e., artificial, composite, inherently 
non-magnetic media with (positive or negative) magnetic response
at microwave frequencies.

Classical MMs are routinely fabricated with regular arrays of split-ring
resonators (SRRs), with operating frequencies up to the optical range.\cite{Yen} 
Moreover, MMs with negative magnetic response can be combined
with plasmonic wires
that exhibit negative permittivity, producing thus left-handed (LH) 
metamaterials characterized by negative refraction index.
Superconducting SRRs promise severe reduction of losses, which constrain
the evanescent wave amplification in these materials.\cite{Ricci}
Thus, metamaterials involving superconducting SRRs and/or wires have 
been recently demonstrated experimentally.\cite{Ricci1}
The effect of incorporating superconductors in LH transmission lines
has been also studied.\cite{Salehi}
Naturally, the theory of metamaterials has been extended to
account for nonlinear effects.
\cite{Zharov-Lapine,Lazarides-Kourakis,Shadrivov,Lazarides1,Kourakis} 
Nonlinear MMs support several types of interesting excitations,
e.g., magnetic domain walls,\cite{Shadrivov} discrete breathers,\cite{Lazarides1}
and evnelope solitons.\cite{Kourakis}
Regular arrays of rf SQUIDs offer an alternative for the construction 
of nonlinear MMs due to the nonlinearity of the JJ.

Very much like the SRR, the rf SQUID (Fig. 1(b)) is a resonant nonlinear oscillator,
and similarly it responds in a manner analogous to a magnetic "atom" 
in a time-varying magnetic field with appropriate polarization,
exhibiting a resonant magnetic response at a particular frequency. 
The SRRs are equivalently RLC circuits in series,
featuring a resistance $R$, a capacitance $C$ and an inductance $L$,
working as small dipoles. 
\begin{figure}[t]
\includegraphics[angle=0, width=.5\linewidth]{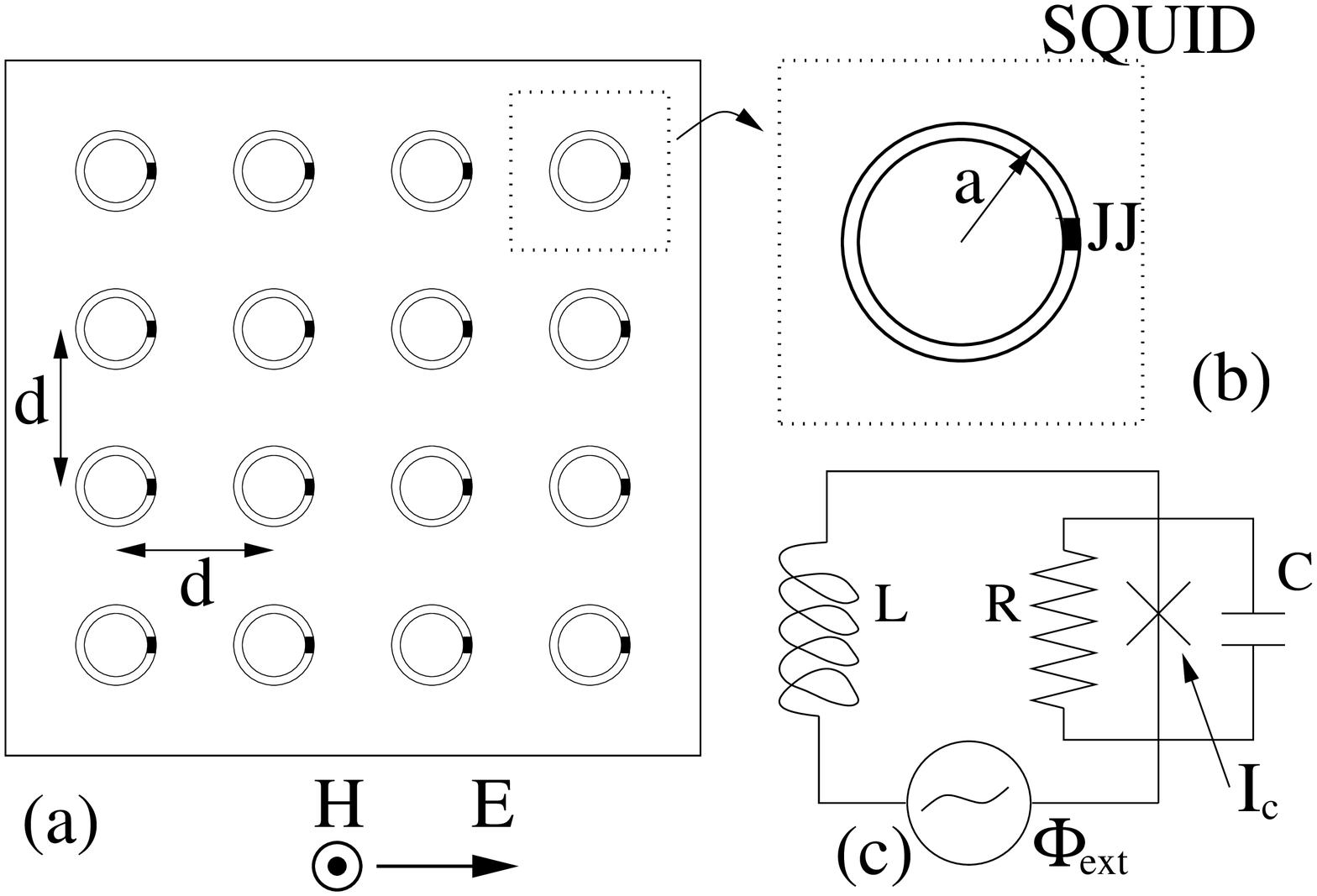}
\caption{
Schematic drawing of the SQUID array, along with the equivalent circuit
for an rf SQUID in external flux $\Phi_{ext}$.  
}
\end{figure}
In turn, adopting the resistively and capacitively shunted junction (RCSJ)
model for the JJ,\cite{Likharev} 
the rf SQUIDs are not dipoles but, instead, they feature an inductance $L$ in series
with an ideal Josephson element (i.e., for which $I=I_c \sin\phi$, with $\phi$
the Josephson phase), 
shunted by a capacitor $C$ and a resistor $R$
(Fig. 1(c)). 
However, the fields they produce are approximatelly those of
small dipoles, although quantitatively they are affected by
flux quantization in superconducting loops. 
Consider an rf SQUID with loop area $S=\pi a^2$ (radius $a$),
in a magnetic field of amplitude $H_{e0}$, frequency $\omega$,
and intensity $H_{ext} = H_{e0} \cos(\omega t)$ 
perpendicular to its plane ($t$ is the time variable).
The field generates a flux $\Phi_{ext} = \Phi_{e0} \cos(\omega t)$
threading the SQUID loop, with $\Phi_{e0} = \mu_0 S H_{e0}$, 
and $\mu_0$ the permeability of the vacuum. 
The flux $\Phi$ trapped in the SQUID ring is given (in normalized variables) by
\begin{eqnarray}
  \label{1}
    f= f_{ext} + \beta\, i ,
\end{eqnarray}
where $f=\Phi/\Phi_0$, $f_{ext} = \Phi_{ext} / \Phi_0$, $i=I/I_c$,
$\beta= \beta_L / 2\pi \equiv L I_c/\Phi_0$,
$I$ is the current circulating in the ring, $I_c$ is the critical current 
of the JJ, $L$ is the inductance of the SQUID ring,
and $\Phi_0$ is the flux quantum.
The dynamics of the normalized flux $f$ is governed by the equation
\begin{eqnarray}
  \label{4}
    \frac{d^2 f}{d\tau^2} +  \gamma \frac{d f}{d\tau} 
    + \beta  \sin\left( 2\pi f \right) + f = f_{ext} ,
\end{eqnarray}
where $C$ and $R$ is the capacitance and resistance, respectively, of the JJ,
$\gamma=L \omega_0 /R$, $\tau=\omega_0 t$, $\omega_0^2=1/LC$, and
\begin{eqnarray}
  \label{5}
   f_{ext} = f_{e0} \cos(\Omega \tau ) , 
\end{eqnarray}   
with $f_{e0} = \Phi_{e0}/\Phi_0$, and $\Omega = \omega / \omega_0$.
The small parameter $\gamma$ actually represents all of the dissipation coupled to
the rf SQUID.

\begin{figure}[t]
\includegraphics[angle=0, width=.65\linewidth]{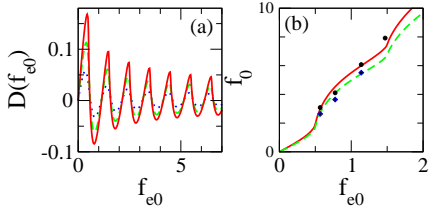}
\caption{(Color online)
(a) Coefficient $D$ vs. the applied flux amplitude $f_{e0}$, 
for $\beta=0.15$ (red-solid curve); $\beta=0.10$ 
(green-dashed curve); $\beta=0.05$ (blue-dotted curve).
(b) The amplitude of the flux $f_0$ vs. $f_{e0}$, 
for $\Omega=0.9$ (red-solid curve), $\Omega=1.1$ (green-dashed curve),
and $\gamma=0.001$, $\beta=0.15$.
The black circles and blue diamonds correspond to the numerically obtained $f_0$
for $\Omega=0.9$ and $1.1$, respectively.
}
\end{figure}
An approximate solution for Eq. (\ref{4}) may be obtained for
$\Omega$ close to the SQUID resonance frequency ($\Omega \sim 1$)
in the non-hysteretic regime $\beta_L < 1$.
Following Ref. \cite{Erne} we expand the nonlinear term in Eq. (\ref{4}) 
in a Fourier - Bessel series of the form
\begin{eqnarray}
 \label{6}
  \beta  \sin\left( 2\pi f \right)
  = -\sum_{n=1}^\infty \frac{(-1)^n}{n\pi} J_n (n\beta_L) \sin(2\pi n f_{ext}) ,
\end{eqnarray}    
where $J_n$ is the Bessel function of the first kind, of order $n$.
By substiting Eq. (\ref{5}) in Eq. (\ref{6}) and carrying out
the Fourier - Bessel expansion of the sine term, one needs to retain only
the fundamental $\Omega$ component in the expansion.\cite{Bulsara}  
This leads to the simplified expression 
\begin{eqnarray}
 \label{7}
    \beta  \sin\left( 2\pi f \right) \simeq D( f_{e0} ) \cos(\Omega \tau) ,
\end{eqnarray}
where $D( f_{e0} )=-2 \sum_{n=1}^\infty \frac{(-1)^n}{n\pi} J_n (n\beta_L) J_1 (2\pi n f_{e0})$.
By substitution of Eq. (\ref{7}) in Eq. (\ref{4}), 
the latter can be solved for the flux $f = f_0 \cos(\Omega \tau +\theta)$
in the loop, with 
\begin{eqnarray}
\label{10} 
 f_0 = \frac{f_{e0} - D}
    {\sqrt{\gamma^2 \Omega^2 + (1-\Omega^2)^2}},
 \qquad \theta = \tan^{-1}\left( \frac{-\gamma\Omega}{1-\Omega^2} \right) ,
\end{eqnarray}  
where $\theta$ is the phase difference between $f$ and $f_{ext}$. 
The dependence of $D$ and $f_0$ on $f_{e0}$
for low field intensity is illustrated in Figs. 1(a) and 1(b), respectively.
For larger $f_{e0}$ the coefficient $D$ approaches zero still oscillating,
while $f_0$ approaches a straight line with slope depending on $\Omega$
and $\gamma$. For $\gamma \ll 1$ and not very close to the resonance, 
$\theta \simeq 0$.
It is instructive to express the $\gamma =0$ solution as: 
\begin{eqnarray}
\label{11} 
f = \pm |f_0| \cos(\Omega \tau), 
\qquad |f_0| = ( f_{e0} - D )/ |1-\Omega^2| . 
\end{eqnarray}  
\begin{figure}[t]
\includegraphics[angle=0, width=.65\linewidth]{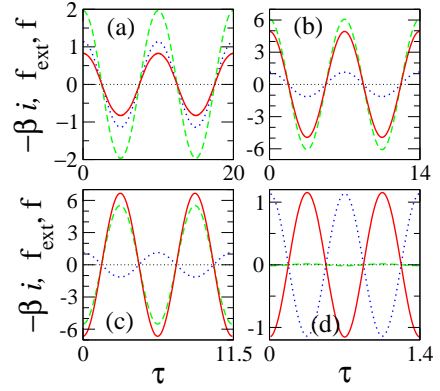}
\caption{(Color online)
Time-dependence of the flux $f$ (green-dashed curves),
the applied flux $f_{ext}$ (blue-dotted curves),
and the response $\beta\, i$ (red-solid curves), 
for $\beta=0.15$, $\gamma=0.001$, $f_{e0}=1.14$, and
(a) $\Omega=0.63$; (b) $\Omega=0.9$; (c) $\Omega=1.1$; (d) $\Omega=9.0$.
}
\end{figure}
The plus (minus) sign, corresponding to a phase-shift of $0$ ($\pi$) 
of $f$ with respect to $f_{ext}$,
is obtained for $\Omega<1$ ($\Omega>1$).
Thus, the flux $f$ may be either in-phase (+ sign) or in anti-phase (- sign)
with $f_{ext}$, depending on $\Omega$. 
This is confirmed by numerical integration of Eq. (\ref{4}), as shown in Fig. 2, 
where we plot separately the three terms of Eq. (\ref{1}) in time.
The quantities $f$, $f_{ext}$, and $\beta\, i$ are shown 
for two periods $T=2\pi/\Omega$ in each case,
after they have reached a steady state. 
For $\Omega<1$
(Figs. 2(a) and 2(b)), the flux $f$ (green-dashed curves) is in-phase
with $f_{ext}$ (blue-dotted curves), while for $\Omega>1$ (Figs. 2(c) and 2(d))
the flux $f$ is in anti-phase with $f_{ext}$.
The other curves (red-solid curves) correspond to $\beta\, i$,
the response of the SQUID to the applied flux. 
Away from the resonance,
the response is (in absolute value) less than (Fig. 2(a), for $\Omega=0.63$) 
or nearly equal (Fig. 2(d), for $\Omega=8.98$) 
to the magnitude of $f_{ext}$.
However, close to resonance, the response $\beta\, i$ is much larger than $f_{ext}$,
leading to a much higher flux $f$ 
(Figs. 2(b) and 2(c) for $\Omega=0.9$ and $\Omega=1.1$, respectively). 
Moreover, in Fig. 2(c),
$f$ is in anti-phase with $f_{ext}$, showing thus extreme diamagnetic (negative)
response.
The numerically obtained amplitudes $f_0$ (depicted as black circles for 
$\Omega=0.9$ and blue diamonds for $\Omega=1.1$ in Fig. 2(b)) are in fair agreement 
with the analytical expression, Eq. (\ref{10}). 
The agreement becomes  better for larger $f_{e0}$.

We now consider a planar rf SQUID array consisting of identical units 
(Fig. 1(a)),
and forming a lattice of unit-cell-side $d$; 
the system is placed 
in a magnetic field $H_{ext}\equiv H$ perpendicular to SQUID plane. 
If the wavelength of $H$ is much larger than $d$,
the array can be treated as an effectivelly continuous and homogeneous
medium. Then, the magnetic induction $B$ in the array plane is 
\begin{eqnarray}
\label{12}
  B = \mu_0 ( H + M ) \equiv \mu_0 \mu_r H , 
\end{eqnarray}
where $M=S\, I/d^3$ is the magnetization induced by the current $I$
circulating a SQUID loop,
and $\mu_r$ the relative permeability of the array.
Introducing $M$ into Eq. (\ref{12}), and using 
Eqs. (\ref{1}), (\ref{5}), and (\ref{11}), we get 
\begin{eqnarray}
\label{13}
 \mu_r = 1 + \tilde{F} \left( \pm |f_0|/{f_{e0}} -1 \right) ,
\end{eqnarray}
where  $\tilde{F} = \pi^2 (\mu_0 a/L) (a/d)^3$.
The coefficient $\tilde{F}$ has to be very small ($\tilde{F} \ll 1$),
so that magnetic interactions between individual SQUIDs can be 
neglected in a first approximation.
Recall that the plus sign in front of $|f_0|/{f_{e0}}$ should be taken
for $\Omega <1$, while the minus sign should be taken for $\Omega >1$.
In Fig. 3 we plot $\mu_r$ both for
$\Omega <1$ (Fig. 3(a)) and $\Omega >1$ (Fig. 3(b)), for three different values 
of $\tilde{F}$.
In real arrays, that coefficient could be engineered to
attain the desired value. 
In both Figs. 3(a) and 3(b), the relative permeability $\mu_r$
oscillates for low intensity fields (low $f_{ext}$), while it tends to 
a constant at larger $f_{ext}$. 
In Fig. 3(a) ($\Omega <1$), the relative permeability $\mu_r$ is always positive, 
while it increases with increasing $\tilde{F}$.
In Fig. 3(b), however, $\mu_r$ may assume both positive and
negative values, depending on the value of $\tilde{F}$. 
With appropriate choise of $\tilde{F}$, it becomes oscillatory around zero
(green-dashed curve in Fig. 3(b))
allowing tuning from positive to negative $\mu_r$  with a slight 
change of $f_{ext}$. 
\begin{figure}[t]
\includegraphics[angle=0, width=.7\linewidth]{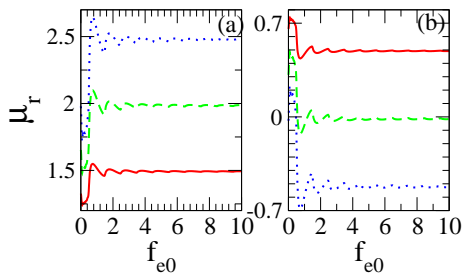}
\caption{(Color online)
Relative permeability $\mu_r$ vs. $f_{e0}$,
for $\tilde{F}=0.01$ (red-solid curves), $\tilde{F}=0.02$ 
(green-dashed curves),  $\tilde{F}=0.03$ (blue-dotted curves),
and $\Omega=0.99$ (a); $1.01$ (b).
}
\end{figure}

In conclusion, we have shown that a planar rf SQUID array exhibits large
magnetic response close to resonance,
which may be negative above the resonance frequency, leading to effectivelly
negative $\mu_r$. 
For low field intensities
(low $f_{ext}$), $\mu_r$ exhibits oscillatory behavior which gradually dissappears
for higher $f_{ext}$. This behavior may be exploited to construct
a flux-controlled metamaterial (as opposed to voltage-controlled metamaterial
demonstrated in Ref. \cite{Reynet}).  
The physical parameters required for the rf SQUIDs giving the dimensionless 
parameters used above are not especially formidable. An rf SQUID with 
$L\simeq 105~pH$, $C \simeq 80~fF$, and $I_c\simeq 3~\mu A$, would give
$\beta \simeq 0.15$ ($\beta_L \simeq 0.94$). 
For these parameters, a value of the resistance 
$R\simeq 3.6 ~K\Omega$
is required in order to have $\gamma \simeq 10^{-3}$, used in the numerical 
integration of Eq. (\ref{4}).  
However, our results are qualitatively valid for $\gamma$ 
even an order of magnitude larger, in which case $R\simeq 360 ~\Omega$.
We note that $\omega_0 = \omega_p/\sqrt{\beta_L}$, where $\omega_p$
is the plasma frequency of the JJ. For the parameters considered above,
where $\beta_L$ is slightly less than unity, the frequencies $\omega_0$
and $\omega_p$ are of the same order. However, $\omega_p$ does not seem
to have any special role in the microwave response of the rf SQUID.
Du {\it et al.} have studied the quantum version of a SQUID array as a 
LH metamaterial, concluded that negative refractive index with low loss 
may be obtained in the quantum regime.\cite{Du}
Consequently, $\mu_r$ can be negative at some specific frequency range.
However, their corresponding expression for $\mu_r$ is linear, i.e.,
it does not depend on the amplitude of the applied flux, and thus
it does not allow flux-tuning. Moreover, experiments with SQUID arrays in the 
quantum regime, where individual SQUIDs can be described as two-level systems,
are much more difficult to realize. 

We acknowledge  support from the grant "Pythagoras II" (KA. 2102/TDY 25)
of the Greek Ministry of Education and the European Union, 
and grant 2006PIV10007 of the Generalitat de Catalunia, Spain.

\end{document}